\documentstyle[12pt,aasms4]{article}

\begin{document}

\title{A Survey of the HCN $J$=1--0 Hyperfine Lines towards Class 0 
and I Sources}

\author{Y.-S. Park\altaffilmark{1,2}, Jongsoo Kim\altaffilmark{1}, 
and Y. C. Minh\altaffilmark{1}} 

\altaffiltext{1}{Korea Astronomy Observatory, San 36-1, Hwaam-dong, Yusung, 
Taejon, 305-348, Korea}
\altaffiltext{2}{Institute of Astronomy and Astrophysics, Academia Sinica, 
P.O. Box 1-87, Nankang, Taipei, Taiwan 115, R.O.C.,
e-mail: yspark@asiaa.sinica.edu.tw} 

\begin{abstract}

The HCN 1--0 hyperfine lines have been observed toward 24 young stellar 
objects (YSOs) of class 0 and I.  The hyperfine lines are well separated 
in most cases and show such rich structures as asymmetric double peaks 
and strong wings.  
We examined how their line shapes and velocity shifts vary along with 
their relative optical depths and compared them with those of CS 2--1, 
H$_2$CO 2$_{12}$--1$_{11}$, and HCO$^+$ 4--3 \& 3--2 transitions 
previously observed by Mardones et al. (1997) and Gregersen et al. (1997).

It is found that all these molecular species do not always exhibit the 
same sense of line asymmetry and the correlation of velocity shift 
is better between HCN and CS than between HCN and H$_2$CO.   
The most opaque transition of HCN $F$=2--1 has about the same velocity 
shift as that of CS despite of the larger beam size of this study, which 
suggests that HCN $F$=2--1 line may be more sensitive to the internal 
motion of YSOs than CS line.  
Systematic changes of the velocity shift are noted for many sources, as 
one goes from $F$=0--1 to 2--1.  The monotonic decrease of velocity 
(blue shift) is apparently more frequent.

A detailed model of radiative transfer allowing line overlap of HCN is 
employed to L483 which shows convincing signatures of infall on a scale 
of $\sim 0.1$ pc.  It appears that the observed line is not compatible 
with the standard Shu (1977) model, but is fitted with augmentations 
of density and infall velocity, by factors of 6 and 0.5, respectively, 
and with an inclusion of a diffuse, static, turbulent, and geometrically 
thick envelope. 

The distribution of hyperfine line intensity ratios for these YSOs does 
not accord with the LTE condition and is essentially the same as ones 
previously noted in cold dark clouds or small translucent cores.  Though 
this anomaly may be explained in terms of radiative transfer effect in 
the cores which are either static or under systematic motion, some of them 
seem to invoke an existence of scattering envelope. 
It is confirmed that HCN is detected more selectively in class 0 and I 
sources than in starless cores or class II objects, which implies that 
the core embedding YSO(s) form a dense ($\sim 10^6$cm$^{-3}$) envelope 
with a significant HCN abundance in a narrow time span of their evolution 
(\cite{afo98}).

\end{abstract}


\section{Introduction}

Since the pioneering works of Zhou and his colleagues (\cite{zho93}; 
\cite{zho94}; \cite{zho95}), observational signatures of infall motion 
in the early phase of star formation have been discovered towards many 
young stellar objects (YSOs).
They are mainly based on the spectral features that an opaque molecular 
line has a self-absorption with the blue peak stronger than the red one 
(hereafter the blue asymmetry), while an optically thin line of single
peak is located between the two peaks of the opaque one.  
In an attempt to search for infalling YSOs, surveys of class 0 and I 
sources have been undertaken using the transitions of, e.g., CS, H$_2$CO, 
and HCO$^+$.  
However, only a few sources like B335 and IRAS 16293--2422 seem to provide 
compelling evidence of collapse motion (\cite{zho93}; \cite{zho95}; 
\cite{nar98}).  Line profiles of some sources exhibit different asymmetry 
from one molecular species to another as well as from transition to 
transition in a molecule (\cite{mar97}; \cite{leh97}; \cite{gre97}).  
It is likely that the infall motion is not so simple or monotonic as one 
expects.  Not only outflow motion but also aspherical geometry make 
the problem more complex.
Thus, previous studies have always come to a conservative conclusion that
class 0 objects seem to undergo infall phase in a statistical sense only 
and it is unclear for class I objects (\cite{mar97}; \cite{gre97}). 

In order to have a comprehensive view on the internal motion of YSOs, one 
needs to make observations with various molecules and transitions using 
both single dish and interferometer.
As a complement to the previous single dish observations by Mardones et al. 
(1997) and by Gregersen et al. (1997), we have carried out an HCN $J$=1--0 
survey of class 0 and I YSOs.  
We can probe denser regions with HCN $J$=1--0 lines than with CS $J$=2--1
one, the most popular density tracer.  The $J$=1--0 transition of HCN has 
three hyperfine lines ($F$=1--0, 1--1, and 2--1) whose optical depths are 
scaled to $1:3:5$ in the LTE condition.  
The hyperfine lines enable us to investigate how the line shape changes 
with their optical depths.  
Moreover, molecular abundance and spatial resolutions are the same for 
the three hyperfine components.  This is an advantage over using other 
combinations of molecules or transitions with different beam sizes or 
abundances.
Recently, it is known that the HCN emission peak of each Bok globule is 
coincident with the position of embedded source within 6$''$ and the 
detection rate of HCN lines is higher in class I and probably class 0 
sources than in starless cores and class II sources (\cite{afo98}).  
Thus HCN may be a good tracer of class 0 and I objects.

The main goal of this study is to provide another set of line profiles 
of YSOs.  Based on the data set, we investigate the tendency of line
asymmetry among three hyperfine components, compare the asymmetry 
with that of other molecular lines, and probe the internal structure 
and kinematics of the cores.  
Probing with various molecular species may also be helpful in 
investigating molecular chemistry in the protostellar cores 
(\cite{raw96}).

After presenting observation and data reduction procedure (section~2),  
we describes observational results in section~3.  A simple radiative
transfer model of L483 is detailed in section~4.  
Discussions on the relevant molecular chemistry and hyperfine line 
ratios of HCN are given in section~5.  
Finally, we summarize results in section~6.

\section{Observations}

The sources listed in Table~1 have been selected based on the data set 
in Mardones et al. (1997) and Gregersen et al. (1997).  We detected 
HCN hyperfine lines toward 22 of 24 with an rms level of a few tens of
mK. 

Observations were carried out with a radome-enclosed 14 meter telescope 
in Taeduk Radio Astronomy Observatory, Korea, during April and May 1997.  
We used an SIS receiver and an autocorrelation spectrometer with 20 KHz 
(0.068 km s$^{-1}$ at HCN $F$=2--1 frequency) resolution.  System 
temperatures were typically $400-600$ K (SSB) and pointing was good to an 
accuracy of 10$''$ in both directions of azimuth and elevation.  
The spectra were Hanning-smoothed, calibrated by a standard chopper wheel 
method, and presented in an antenna temperature ($T_A^*$) scale.  

The rest frequency of $F$=2--1 transition is 88.631847 GHz, and the 
separations of $F$=1--1 and $F$=0--1 lines with respect to $F$=2--1 are 
4.84 and $-7.07$ km s$^{-1}$, respectively (\cite{lov92}).  
Measured FWHM beam size and beam efficiency are 61$''$ and 40\%, 
respectively, at this frequency.  
We used frequency switching mode with $\Delta f = 6$ MHz and obtained the 
spectra with the S/N ratios better than 15.

\section{Results}

\subsection{Line profiles}

Observed line profiles are presented in Fig.~1 (a-d), where we see 
spectral features like self-absorption, asymmetry, and broad wings.  
The hyperfine lines are well separated except for a few cases.  Line 
asymmetries of many sources seem to grow as the optical depth increases.  
All three hyperfine lines appear to be optically thin for L146, Serp SMM3, 
L673A, and L1152, whereas they all are optically thick for NGC1333-4A, 
L483, and B335.  
The line wings of many sources are generally weaker than those in other 
studies (\cite{mar97}; \cite{gre97}), which may result from a larger 
telescope beam of this study.

\subsection{Individual sources}

We outline the characteristics of HCN line profiles of the sources and, 
when available, compare them with those of CS 2--1, H$_2$CO 2$_{12}$--1$_{11}$, 
and N$_2$H$^+$ 1--0 in Mardones et al. (1997) and with HCO$^+$ 4--3 
and 3--2 in Gregersen et al. (1997).
Mainly sources with dissimilarities in line shapes among molecular 
species are discussed.
It should be noted that the two surveys are made with FWHM beam
size of $\stackrel{<}{_{\sim}}20''$, while ours are with 61$''$.

\noindent 
L1448-IRS3: $F$=2--1 line looks similar to CS line, whereas $F$=0--1 line 
to N$_2$H$^+$ line.  HCO$^+$ 3--2 line is significantly shifted to the red.

\noindent 
L1448mm: All HCN hyperfine components show the blue asymmetry like CS 
spectrum.  On the other hand, all transitions of H$_2$CO and HCO$^+$ have 
the red asymmetry.

\noindent 
NGC1333-2: All three lines differ from those of CS and H$_2$CO in their 
shapes.  Instead, they are similar to N$_2$H$^+$ line, which seems to be 
also optically thick.

\noindent 
L43: It shows how the absorption dip grows as the optical depth increases.  
The line shape of $F$=2--1 transition is quite similar to that of CS.  
$F$=0--1 line is relatively strong compared to the other two hyperfine 
lines.

\noindent 
L146: The shape of line profiles are closer to that of H$_2$CO.  Lines are 
gradually shifted to the red, as the transition becomes optically thick.  
As in L43, $F$=0--1 line is brighter than $F$=1--1 line.  $F$=2--1 line is 
quite narrow.

\noindent 
L483: This is a good example showing how the line shape varies with the 
optical depth.  It seems to be consistent with a Shu (1977) type infall 
motion; the red shoulder in the least opaque $F$=0--1 transition changes 
into the absorption dip in the most opaque $F$=2--1 one.  
$F$=2--1 lines are more asymmetric than those of CS and H$_2$CO.  However,
the asymmetry is reversed to the red in the 20$''$ beam observation of 
HCO$^+$.  It is interesting to note that $F$=0--1 line is the brightest 
among three components.  The depression of $F$=2--1 line can not be 
explained by self-absorption only (see section~4).

\noindent 
S68N: $F$=2--1 line has a deep self-absorption with a wider blue component.  
The general shape of HCN resembles that of CS, though the blue wing of HCN 
is weaker.  By contrast, H$_2$CO has a clear blue asymmetry.

\noindent 
FIRS1: This is a typical object demonstrating diverse asymmetries of  
different molecular transitions.  $F$=2--1 component of HCN shows a 
prominent blue asymmetry like CS, H$_2$CO line does the red one, and 
HCO$^+$ lines are nearly symmetric. 

\noindent 
SMM4: This shows a progressive blue shift of HCN lines and deviation 
from symmetry with an increasing optical depth.  However, the shift 
may not result from infall, but from outflow motion, judged from long 
tail to the blue. 
The opaque lines are similar in shape to both CS and H$_2$CO lines.
HCO$^+$ lines with the central absorption dips have the same asymmetry, 
but do not have any long tail to the blue. 

\noindent 
SMM3: All three HCN components are fitted well with Gaussian.  Other 
transitions of CS, H$_2$CO, and HCO$^+$ are also symmetric.  Relative 
intensities are quite close to the values of optically thin limit under 
the LTE condition.

\noindent 
B335: The asymmetry of HCN lines seems to change from the blue to the red 
as we move from $F$=0--1 line to $F$=2--1 one, although the S/N ratio is 
not so high.  
B335 core has been well known as a prototype of YSOs with convincing 
evidence of core collapse, and thus well studied (\cite{zho93}, 1994; 
\cite{zho95}; \cite{cho95}).  However, our observation contradicts most 
existing observations (cf. \cite{kam85}), except CCS observations 
(\cite{vel95}).  DC 303.8-14.2 is another example showing a reversal 
of asymmetry among HCN hyperfine lines (\cite{leh97}).  Most observations 
suggestive of collapse motion have been made with a beam of 
$\stackrel{<}{_{\sim}} 20''$, while observations of this study and 
Velusamy et al. (1995) are with $\sim 1'$ beam.  Line profiles obtained 
with larger beams may be contaminated by outflow motion.  
However, we can not rule out a possibility that the outer part of B335 is 
slowly expanding (\cite{leh97}).  

\noindent 
L1157: $F$=2--1 and 1--1 lines have long tails to the blue with single 
peaks, while $F$=0--1 line is almost symmetric.  Both CS and H$_2$CO lines 
show the blue asymmetry as well.  On the contrary, HCO$^+$ lines have 
rather strange spectral features, the blue asymmetry in $J$=3--2 and the 
red one in $J$=4--3 transition.  

\noindent 
L1251B: Shoulders in the red of $F$=2--1 and 1--1 transitions coincide 
with that of CS.  H$_2$CO has the blue asymmetry like HCN and CS, but its 
self-absorption is much deeper than those of HCN and CS.  

In summary, for a large fraction of sources, the asymmetries are the same 
in all the transitions of HCN, CS, H$_2$CO, and HCO$^+$.  However, for 
several sources, it depends on the molecular species and, particularly 
for B335 and L1157, it varies from transition to transition of a molecule.
It should be pointed out that the asymmetry of HCN is relatively more 
similar to that of CS than to that of H$_2$CO, which will be quantified 
in next subsection.

\subsection{Correlations}
 
For more detailed investigation of line profiles, one needs to quantify 
the line velocity associated with internal motion.  We use a parameter 
defined by Mardones et al. (1997).  
At first, we measure $V_{\rm G}$($i$--$j$), the Gaussian peak velocity 
in the $F$=$i$--$j$ transition of HCN after fitting the three hyperfine 
lines with three Gaussians.
Fitting procedure is straightforward for the lines of simple shape. 
For lines having another weak peak or a shoulder, we fit the profile after 
masking such features.  If intensities of two peaks differ by less than 
$2 \sigma$, where $\sigma$ is the rms noise of spectrum, the whole velocity 
span is taken into account.
The resulting $V_{\rm G}$($i$--$j$) and its standard deviation of each 
hyperfine component are listed in Table~1.  The standard deviation is 
usually less than one channel width of the spectrometer.  
Since some hyperfine transitions of S68N, Serp FIRS1, and Serp SMM5 are 
blended each other, their $V_{\rm G}$($i$--$j$)s are not included in the 
table. 
The measure of velocity shift or degree of asymmetry $\delta V$ is 
then defined as
\begin{equation}
\delta V = (V_{\rm G} - V_{\rm thin})/\Delta V_{\rm thin}.
\end{equation}
We use the line velocity and FWHM of N$_2$H$^+$ for $V_{\rm thin}$ 
and $\Delta V_{\rm thin}$, respectively, in Mardones et al. (1997).

We plot relations between $\delta V$(HCN) and $\delta V$(CS), and between 
$\delta V$(HCN) and $\delta V$(H$_2$CO) in Fig.~2.
The velocity shifts of CS and H$_2$CO molecules are also from Mardones 
et al. (1997).  It is found that $\delta V$(0--1) correlates well with 
$\delta V$(CS).  And we do not find any significant differences in their 
relations between class 0 and I.  The close relation itself implies that 
HCN $F$=0--1 transition probes as similar kinematics as CS molecule does.  
It would be possible only if the distribution of HCN molecule is similar 
to that of CS inside the core.  
However, the slope of 0.47 suggests that $F$=0--1 transition is moderately 
optically thick; if the transition is as thin as that of N$_2$H$^+$, the 
slope should be around zero.  
It would not be so optically thick, since their line shapes are usually 
simple and show a single peak.  The slope significantly less than unity 
may also be due to a difference in the beam size of two surveys; our 
larger telescope beam, which covers more volume of static envelope, 
may lessen the slope.  

As going toward the $F$=2--1 transition, we notice the worse correlation
as well as the gradual increase of the slope.
The increase of the slope up to around unity implys that the most opaque 
line of HCN is as sensitive to the internal motion of the core as the CS, 
again suggesting similar spatial distributions of both HCN and CS molecules.  
General similarity in the line shapes between HCN $F$=2--1 and CS $J$=2--1, 
as shown in the previous subsection, supports this argument.  An increasing 
scatter can be interpreted in terms of the opacity of the transition; 
since the opaque line is formed where the optical depth is about one, 
it does not reflect a global property, but does a local one.  Thus small 
differences in excitation conditions and chemistry between HCN and CS would
result in large differences in their line profiles.

In the lower panel of Fig.~2, we plot the relations between $\delta V$(HCN) 
and $\delta V$(H$_2$CO).  The correlation between them is not so good as 
the case of $\delta V$(HCN) and $\delta V$(CS); we can hardly find a 
linear relationship between $\delta V$(0--1) and $\delta V$(H$_2$CO). 
Their distributions mimic the relation between $\delta V$(CS) and 
$\delta V$(H$_2$CO) illustrated in Fig~3, which is drawn from data set
of Mardones et al. (1997). 
However, if we confine ourselves to the sources of $\delta V$(HCN)$<0$ 
and $\delta V$(H$_2$CO)$<0$, then we can see as similar trend 
as in the upper panel -- the concomitant increase of the slope to unity 
with the optical depth.  

Good correlations of $\delta V$s between HCN and CS indicate a similar
distribution of both molecular species in the core.  It is intriguing 
that, in spite of the larger beam size of our HCN observation, the slope 
is roughly unity for both pairs of $\delta V$(HCN)--$\delta V$(CS) and 
$\delta V$(HCN)--$\delta V$(H$_2$CO) for the most opaque transition of 
HCN.  It appears that HCN is even more sensitive to the internal motion 
of the core.
The critical density of HCN $J$=1--0, $2.5\times 10^6$ cm$^{-3}$, 
is larger than those of CS and H$_2$CO, $5.7\times 10^5$ cm$^{-3}$ and  
$1.1\times 10^6$ cm$^{-3}$, respectively.
Thus HCN molecule traces deeper and denser regions of YSOs than CS or 
H$_2$CO.  If HCN lines are observed with as similar beam size as that 
of CS or H$_2$CO ($\sim 20''$), we may be able to see more clearly 
the internal motion of the cores. 

Fig.~4 quantifies the $\delta V$(HCN) of individual sources as a function 
of relative optical depth.  If there is any systematic internal motion,
we may expect gradual increase or decrease of $\delta V$ (\cite{mye95}; 
\cite{zha98}). 
It was not so easy to find such a relation from combinations of different 
molecules or transitions often sparsely distributed in frequency space, 
since the chemistry and beam size are different from each other.  
Most YSOs exhibit a monotonic increase or decrease of $\delta V$ with 
respect to the optical depth.  Seven sources (L1448mm, NGC1333-4A, L483, 
Serp SMM4, L1157, L1172, and L1251B) suggest infall motion, while three 
sources (L43, L146, and B335) do outward motion.  
A few YSOs (L1448-IRS3, serp SMM3, and 18331-0035) seem to be static.  
A few large blobs or clumps may also give rise to such a systematic 
velocity shift.  This possibility is, however, ruled out by the 
smoothness of optically thin N$_2$H$^+$ lines which are almost symmetric.

\section{Radiative Transfer Model of L483}

Line profiles of L483 in Fig.~1 demonstrate clearly how the absorption 
dip develops under inward velocity field as the optical depth increases.
Line intensity ratios among hyperfine transitions significantly deviate 
from the `standard hyperfine ratios' (see section~5).  
Thus, synthesizing them with radiative transfer code will be useful in 
understanding the infall motion as well as structure of L483.  
We do not try to reproduce detailed features of the observed line profiles, 
but focus on quantitative comparisons between observed and synthesized 
lines. 

The radiative transfer of HCN is complicated due to line overlap 
caused by hyperfine splitting of energy levels.  In the case of cold 
cores like L483, the hyperfine lines do not overlap each other in 
$J$=1--0 transition, but they do in the transitions of higher $J$s, 
which affects the excitation condition of $J$=1--0 transition.  The 
problem of the line overlap has been successfully treated by the Monte 
Carlo method (\cite{gon93}; \cite{lap89}) as well as by a conventional 
one (\cite{tur97}).  We will use the former scheme and impose the condition 
of infall motion.  Levels up to $J$=4 are taken into account, which 
include 13 individual energy levels and 21 radiative transitions among 
them.  The Einstein $A$ coefficients and line frequencies are provided 
by Gonz\'{a}les-Alfonso (1998) and by Turner (1998), and the collisional
rate coefficients by Monteiro \& Stutzki (1986).  With those molecular 
constants, we made a model of one dimensional radiative transfer, and 
confirmed its performance, by reproducing results of Gonz\'{a}les-Alfonso 
\& Cernicharo (1993).

The Shu model (\cite{shu77}) could be a start point for the distributions 
of gas density and motion inside the core.  Based on the observation of 
high density tracing molecules, HC$_3$N and NH$_3$ (\cite{ful93}), we fix 
the size of the core as $1.\!\!'72$ or $R=0.10$ pc at a distance 
of 200 pc.
The core is divided into 30 concentric shells with radii running as 
$r_i \propto i^{0.7} (i=0,...,30)$.  Since temperature is found to be 
12 K near the {\it IRAS} source and fall to 9 K in the outer region 
(\cite{ful93}; \cite{par91}), $T_k$ is assumed to be constant at 10 K 
in the first set of calculations.  The sound speed is then 0.2 km s$^{-1}$. 
An e-folding non-thermal turbulence of $v_{\rm turb}=0.4$ km s$^{-1}$ 
is found from Myers et al. (1995), and the abundance of HCN relative to 
H$_2$, $5\times 10^{-9}$ is from Turner et al. (1997) and from 
Gonz\'{a}les-Alfonso \& Cernicharo (1993).  
Model calculations are then carried out and resulting line profiles are 
convolved with the telescope beam. 
In order to compare the synthetic line profile with the observed one 
in brightness temperature unit, we divided the latter one by 0.5, a 
compromise between the main beam efficiency (0.4) and the forward beam 
coupling efficiency (0.7).

The Shu model is completely described by an infall radius, $r_{\rm inf}$ 
or an elapsed time after the onset of infall.  Fig.~5 shows synthesized 
line profiles for $r_{\rm inf} = 0, 0.4 R$, and $0.8 R$, respectively. 
From the synthesized lines (top in Fig.~5), we find that
i) the asymmetry is negligible in three hyperfine lines, 
ii) the synthesized lines are weaker than the observed ones,
and iii) the line intensity ratios are maintained as 
$I(F=0-1):I(F=1-1):I(F=2-1) = 1:(1.3-2):(2-3)$.  Then we applied the 
kinetic temperature varying as $T_k(r)=10 (R/r)^{0.4}$ K, where the 
$r^{-0.4}$ dependence is based on the distribution of dust temperature 
(\cite{mye95}; \cite{wan95}).
The sound speed was accordingly changed to 0.25 km s$^{-1}$.  In fact,
the Shu model assumes `isothermal' cloud, and thus the model with 
kinetic temperature decreasing outward is far from consistency.  
However, we need to assume `constant' sound speed in order to use his 
simple expression of density and velocity field.  With this 
new temperature distribution, the synthetic lines (middle in Fig.~5) 
become stronger and more asymmetric.  However, the $F$=0--1 line is 
still weaker than the observed one, and the hyperfine ratios differ from 
observation.  
It appears that the inside-out collapse model can not produce the 
observational features well.
One viable option is to increase the density of Shu model core and to 
introduce an extended diffuse envelope, as proposed by Wang et al. 
(1995) and by Gonz\'{a}les-Alfonso \& Cernicharo (1993).
The role of the diffuse envelope is to attenuate the optically thicker 
line more.  A large turbulence in the envelope is required for 
rather uniform attenuation across the line. 
If the envelope is sufficiently opaque, the line core of $F$=2--1 
component may be formed in the envelope.  The general weakening of 
lines is then compensated by an augmentation of density in the core.  
A successful fit after several trial and errors is shown in the 
bottom of Fig.~5.  The resulting model is that a core with 
$r_{\rm inf}=R=0.1$ pc and $v_{\rm turb}=0.3$ km s$^{-1}$ is embedded in 
a static envelope with $R=0.3$ pc, $n({\rm H}_2)=5\times 10^3$ cm$^{-3}$, 
$v_{\rm turb}=1$ km s$^{-1}$, and $T_k=10$ K.  In the core, the density  
increases by a factor of 6 and the infall velocity decreases by a factor 
of 2 with respect to those of the Shu model, respectively.  
The line center optical depths of $F$=0--1, 1--1, and 2--1 transitions
towards the center of core are 3, 8, and 11, respectively, which are not 
beam averaged.  The difference in peak brightness temperature is, of 
course, due to different excitation temperatures of the transitions.
The red shift of absorption dip shown in Fig.~1b justifies rather large
infall radius; without the inward motion of outer layer, the absorption 
dip will be located at the velocity of optically thin line.
The size and density of envelope may change in such a way that its 
optical depth is kept constant.  Different combinations, however, give 
rise to deeper or shallower absorption dips due to different excitation 
conditions of HCN in the envelope.  
Thus, we come to a model different from that of Shu (1977), as invoked 
already by Wang et al. (1995).  In fact, higher density and slower infall 
speed are characteristic features of a magnetically supported core model, 
where the contraction occurs quasi-statically and the density of envelope 
increases as the core evolves (\cite{cio94}, 1995).

Obviously, the real structure of L483 seems to be more complex than our 
model.  It is shown that a near IR image exhibits an elongated structure 
in the East-West direction and the axis of CO molecular outflow is also 
aligned in this direction (\cite{par91}; \cite{lad91}; \cite{ful95}).  
Gregersen et al. (1997) noticed that HCO$^+$ $J$=3--2 and 4--3 lines 
show the red asymmetry, contrary to ours.  However, the possibility of 
outward motion suggested by HCO$^+$ near the center is ruled out by the 
recent VLA observation of NH$_3$ indicating inward motion down to 
$\sim 0.\!\!'2$ scale (\cite{ful99}).  
Reasons for different asymmetry of HCO$^+$ may be attributable to 
differences in molecular chemistry and excitation conditions (see 
section~5).

\section{Discussions}


The similarity of line shapes between CS and HCN and the dissimilarity
between HCO$^+$ and HCN were already mentioned in section~3.  We further 
examine the line asymmetries of 9 class 0 objects which have been 
observed in all transitions of four molecular species (CS, H$_2$CO, HCN, 
and HCO$^+$).
As summarized in Table~2, only one source each shows the reversal of 
asymmetry between CS and HCN, and between HCO$^+$ and H$_2$CO.  
On the other hand, three sources change their asymmetries between HCN 
and HCO$^+$.  Thus, though the number of samples is not large, it seems 
that the line asymmetry is similar within each pair of CS--HCN and 
HCO$^+$--H$_2$CO, but different between two pairs. 
It is interesting to note that CS and probably HCN seem to prefer the
blue asymmetry compared with HCO$^+$.

Why do the two pairs show different asymmetry?  Let us first consider 
if the critical densities of four molecules are grouped in the same way 
as their line shapes.  A sequence of transitions is, however, 
CS 2--1, H$_2$CO 2$_{12}$--1$_{11}$, HCN 1--0, and HCO$^+$ 3--2 \& 
4--3 in an order of increasing critical density (The critical densities 
of HCO$^+$ 3--2 and 4--3 are $0.5\times 10^7$ and $1.3\times 10^7$ 
cm $^{-3}$, respectively; \cite{mon85}), suggesting that excitation 
condition is of little importance.
Then molecular chemistry such as depletion in the innermost region and 
enhancement in the outflow may play important roles in the line formation.  
It is known that HCN as well as CS may freeze out onto grains in cold
dense region (\cite{ber95}; \cite{bla92}; \cite{mcm94}), whereas HCO$^+$ 
will not, since it has a small dipole moment and there is no chemical 
reaction route to consume it in this region (\cite{raw96}; \cite{van98}).  
However, there seems to be little observational evidence of significantly 
depleted CS or HCN.  
If so, the blue asymmetry would be more frequent in HCO$^+$ than in CS 
or HCN, provided that motion is more like inside-out collapse, which is 
usually confined to the central region of core.
However, this is not the case, as shown in Table~2.


Outflows prevalent in these sources may make certain molecules more 
abundant.  L1157 is one of the examples which shows dramatic enhancements 
of various kinds of molecules (\cite{bac97}), where four molecules of our 
concern, CS, H$_2$CO, HCN, and HCO$^+$ are all enhanced.
In the interferometric observations of 9 class 0 objects, HCO$^+$ and 
HCN show so different sensitivity to the envelope and outflow from source 
to source that one could not find any general tendency (\cite{cho99}).
Recent BIMA observation of L483 suggests that HCO$^+$ traces outflow, while 
HCN does thick disk or envelope (\cite{par99}).
The molecular chemistry is very complicated in this way, if both outflow 
and infalling envelope coexist; even in the case that the outflow occupies 
a small fraction of volume, the line shape will be significantly affected, 
if the abundance increases drastically (e.g., $>100$) in the flow region.
The chemistry may also be related with the evolution of YSOs.  Observations 
with high spatial resolution and more elaborated time-dependent chemistry 
models are required for further understanding of line formation and 
molecular chemistry in YSOs (\cite{mun95}).

%
%


We noted in section~3 that hyperfine line intensities of Serp SMM4 
and SMM3 are scaled to the relative LTE optical depth. 
It is also interesting to note that, despite of their strong 
self-absorptions, the intensity ratios of three sources, S68N, 
Serp FIRS1, and SMM5, are also close to $1:3:5$, an optically thin 
limit in the LTE condition. 
In Fig.~6, excluding these three sources, we plot the hyperfine line 
ratios, $R_{02}$ and $R_{12}$, defined by,
\begin{eqnarray}
R_{02} = {T_{\rm max}(F=0-1) \over T_{\rm max}(F=2-1)}, \ \ \ \ \
R_{12} = {T_{\rm max}(F=1-1) \over T_{\rm max}(F=2-1)}. 
\end{eqnarray}
It is found that $R_{12}$ lies between 0.4 and 0.7, except one source, 
while $R_{02}$ spans from 0.2 to 1.0.  Similar distributions have been
known for quiescent dark clouds (\cite{har89}; \cite{gon93}) and for 
small translucent cores (\cite{tur97}), though their physical conditions 
are quite different each other.  The parameter space with 
$R_{12}\stackrel{<}{_{\sim}}1$ and $R_{02}\stackrel{<}{_{\sim}}0.8$ can 
be explained in terms of radiative transfer effect on the clouds/cores 
which are either static or under systematic motion (\cite{gon93}; 
\cite{tur97}).  However, the region of 
$R_{12}\stackrel{>}{_{\sim}}1$ or $R_{02}\stackrel{>}{_{\sim}}0.8$, 
where the $F$=2--1 component is significantly suppressed with respect 
to the $F$=0--1 and $F$=1--1 components, seems to invoke the existence 
of diffuse envelope, as we have shown in section~4.


Recently, Afonso et al. (1998) carried out an HCN $J$=1--0 survey 
towards YSOs in Bok globules.  They found that HCN is detected with a 
higher probability in class I and probably class 0 objects than in 
starless cores and class II sources.  Because there is only one class 0 
source in their survey, such a preference for class 0 was uncertain.
Our study implies the high detection rate of almost unity for both class 0 
and I sources.  Thus there seems to be a unique phase of class 0 and I  
in the evolution of YSOs when a dense envelope of $\sim 1'$ ($0.05-0.1$
pc at the distances of $150-300$ pc) in size is formed.  

\section{Summary}

We have carried out a survey of HCN $J$=1--0 hyperfine lines for 24
objects identified as class 0 and I with a spectral resolution of 
0.068 km s$^{-1}$.  22 sources are detected with around 30 mK rms level 
in $T_A^*$ unit.

It is found that three hyperfine components show a variety of spectral 
features such as deep self-absorption, asymmetry, and broad wings which  
are more prominent in the optically thicker lines.  
Moreover, for a large fraction of sources, HCN hyperfine lines 
show a progressive shift to the blue, as the optical depth increases.  
Only a few sources show a gradual shift to the red, which implies that 
an inward motion is predominant in the core embedding YSOs.
When compared with previous CS and H$_2$CO surveys, the velocity shifts 
of HCN correlate better with those of CS than with those of H$_2$CO.
Little difference in the correlation is noted between class 0 and I.

L483 is confirmed as a candidate infalling source on $\sim 0.1$ pc scale, 
from a growing degree of asymmetry and self-absorption with an increasing 
optical depth.
We synthesized its hyperfine lines, by solving radiative transfer in a 
collapsing core model with the Monte Carlo method.
The synthetic lines based on the Shu (1977) model do not fit the observed 
ones.  We reproduced the observed ones successfully with the modification 
of the Shu model, an overall increase of gas density by a factor of 6 and 
the decrease of infall velocity by a factor of 2.  
Furthermore, in order to explain the line intensity ratios, a diffuse, 
static, and geometrically thick envelope surrounding the modified Shu core 
is essential.

\acknowledgments

Authors are grateful to Dr. C.W. Lee and Dr. E. Gonz\'{a}les-Alfonso for 
helpful discussions.  This study was supported by Korea Astronomy 
Observatory through KAO grant 97-5400-000.
 

\clearpage

\renewcommand{\baselinestretch}{1.2}

\figcaption{{\it a-d.} The observed line profiles of HCN.  Three hyperfine 
components of $F$=0--1 (bottom), $F$=1--1 (middle), and $F$=2--1 (top) are 
aligned in order to see any systematic changes in their line shapes and 
positions with optical depth.
In each panel, a vertical line marks the LSR velocity determined by the 
optically thin N$_2$H$^+$ 1--0 transition (Mardones et al. 1997).  For 
Serp SMM2, HC$^{18}$O$^+$ 3--2 line is used instead (Gregersen et al. 1997).  
 \label{fig1}}

\figcaption{The relation between $\delta V$(HCN) and $\delta V$(CS) and 
between $\delta V$(HCN) and $\delta V$(H$_2$CO), where filled circles are 
for class 0, and filled squares for class I.  
We derive linear relations (solid lines) of the form $y=ax$ by least 
square fit, assuming noise in {\it both} variables.  In the top right are 
linear correlation coefficients.  Error bars of $\pm 1 \sigma$ in the 
lower right corners are the largest ones among these sources. 
 \label{fig2}}

\figcaption{The relation between $\delta V$(CS) and $\delta V$(H$_2$CO). 
Data are from Mardones et al. (1997).  Circles and squares represent 
class 0 and I, respectively.  Same as in Fig.~2 for the error bars.
\label{fig3}}

\figcaption{The $\delta V$(HCN) as a function of the relative optical 
depth.  The optical depth of N$_2$H$^+$ is assumed to be zero. 
Error bars of $\pm 1 \sigma$ are indicated for each source.
Vertical offsets are arbitrary.
 \label{fig4}}

\figcaption{The comparison of synthetic (lines) and observed (histogram)
line profiles of L483.  Top three profiles are from the standard Shu model 
with a constant $T_k$ of 10 K.  Solid, dotted, and dashed lines correspond 
to $r_{\rm inf}=0, 0.4 R$, and $0.8 R$, respectively.
Three profiles in the middle are from the standard Shu model with a
temperature variation of $T_k=10 (r/R)^{-0.4}$ K.  Notations are the same 
as above.  
Two in the bottom are the profiles of the best fit and the observed ones.  
See text for model parameters of the best fit. 
 \label{fig5}}

\figcaption{The distribution of intensity ratios of HCN hyperfine lines, 
$R_{12}$ and $R_{02}$ (see text for definition). 
A solid line represents a locus under the LTE condition; the $R_{12}$ and 
$R_{02}$ approach to unity, as the transition becomes opaque.
 \label{fig6}}

\clearpage

\begin{deluxetable}{lrrccrrr}
\footnotesize
\tablecolumns{8}
\tablecaption{
List of sources and their line velocities
 \label{tbl1}}
\tablehead{
\colhead{Name} & \colhead{RA(1950)} & \colhead{\ \ DEC(1950)} & 
\colhead{class$^{\rm a}$}
              & \colhead{}    & \multicolumn{3}{c}{$V_{\rm G}$(km s$^{-1}$)} \\
\colhead{} &  \colhead{$h \ m \ s$} 
& \colhead{$^{\circ}$ \ $'$ \ $'$} \\
\cline{6-8}   \\
\colhead{}  & \colhead{} & \colhead{} & \colhead{} & \colhead{}
              & \colhead{$F$=0--1} & \colhead{$F$=1--1} & \colhead{$F$=2--1} 
}
\startdata
L1448-IRS3 &  3 22 31.5 &  30 34 49 & 0 & & 4.53$\pm0.02$ & 4.52$\pm0.02$ & 
4.54$\pm0.01$ \nl
L1448mm    &  3 22 34.4 &  30 33 35 & 0 & & 4.76$\pm0.03$ & 4.84$\pm0.02$ & 
4.82$\pm0.02$ \nl
NGC1333-2  &  3 25 49.9 &  31 04 16 & 0 & & 7.65$\pm0.04$ & 7.59$\pm0.03$ & 
7.50$\pm0.02$ \nl
NGC1333-4A &  3 26 04.8 &  31 03 13 & 0 & & 7.04$\pm0.04$ & 6.76$\pm0.02$ & 
6.70$\pm0.01$ \nl
L43        & 16 31 37.7 & -15 40 52 & I & & 0.58$\pm0.01$ & 0.54$\pm0.02$ & 
0.71$\pm0.01$ \nl
L146       & 16 54 27.2 & -16 04 48 & I & & 5.25$\pm0.01$ & 5.24$\pm0.02$ & 
5.34$\pm0.01$ \nl
L483       & 18 14 50.6 & -04 40 49 & 0 & & 5.30$\pm0.01$ & 5.11$\pm0.03$ & 
5.05$\pm0.02$ \nl
S68N       & 18 27 15.2 &  01 14 57 & 0 & & \nodata & \nodata & \nodata \nl
Serp FIRS1 & 18 27 17.4 &  01 13 16 & 0 & & \nodata & \nodata & \nodata \nl
Serp SMM5  & 18 27 18.9 &  01 14 36 & 0 & & \nodata & \nodata & \nodata \nl
Serp SMM4  & 18 27 24.3 &  01 11 11 & 0 & & 7.81$\pm0.03$ & 7.62$\pm0.02$ & 
7.71$\pm0.01$ \nl
Serp SMM3  & 18 27 27.3 &  01 11 55 & 0 & & 7.80$\pm0.03$ & 7.70$\pm0.01$ & 
7.68$\pm0.01$ \nl
Serp SMM2  & 18 27 28.0 &  01 10 45 & 0 & & 7.45$\pm0.03$ & 7.30$\pm0.02$ & 
7.28$\pm0.01$ \nl
18331-0035 & 18 33 07.6 & -00 35 48 & 0 & &10.86$\pm0.03$ & 10.81$\pm0.02$ & 
10.89$\pm0.01$ \nl
L723       & 19 15 41.3 &  19 06 47 & 0 & & \nodata & \nodata & \nodata \nl                                 
L673A      & 19 18 04.6 &  11 14 12 & 0 & & 7.02$\pm0.02$ & 6.94$\pm0.01$ & 
6.99$\pm0.01$ \nl
B335       & 19 34 35.7 &  07 27 20 & 0 & & 8.38$\pm0.02$ & 8.39$\pm0.02$ & 
8.60$\pm0.01$ \nl
IRAS20050  & 20 05 02.5 &  27 20 09 & 0 & & \nodata & \nodata & \nodata \nl 
L1152      & 20 35 19.4 &  67 42 30 & I & & 2.67$\pm0.01$ & 2.65$\pm0.02$ & 
2.62$\pm0.01$ \nl
L1157      & 20 38 39.6 &  67 51 33 & 0 & & 2.60$\pm0.02$ & 2.38$\pm0.02$ & 
2.31$\pm0.01$ \nl
L1172      & 21 01 44.2 &  67 42 24 & I & & 2.84$\pm0.02$ & 2.60$\pm0.02$ & 
2.57$\pm0.02$ \nl
L1251A     & 22 34 22.0 &  75 01 32 & I & &-4.90$\pm0.04$ & -4.98$\pm0.04$ & 
-5.15$\pm0.02$ \nl
L1251B     & 22 37 40.8 &  74 55 50 & I & &-3.94$\pm0.03$ & -4.06$\pm0.03$ & 
-4.13$\pm0.01$ \nl
L1262      & 23 23 48.7 &  74 01 08 & I & & 4.12$\pm0.02$ & 4.08$\pm0.03$ & 
4.25$\pm0.02$ \nl

\enddata
\tablenotetext{}{$^{\rm a}$ Sources are classified as class 0, if 
$T_{\rm bol}\leq 80$ K, and class I, otherwise.  } 
\end{deluxetable}

\clearpage

\begin{deluxetable}{lccccc}
\footnotesize
\tablecolumns{6}
\tablecaption{
Line asymmetry
 \label{tbl2}}
\tablehead{
\colhead{Name} & \colhead{CS 2--1} & \colhead{H$_2$CO 2$_{12}$--1$_{11}$} &
\colhead{HCO$^+$ 4--3} & \colhead{HCO$^+$ 3--2} & \colhead{HCN $F$=2--1} 
}
\startdata
L1448-IRS3 & R & R & \nodata & R & N \nl
L1448mm    & B & R & R       & R & B \nl 
NGC1333-4A & B & B & B       & B & B \nl
L483       & B & B & R       & R & B \nl
Serp FIRS1 & B & R & N       & N & B \nl
Serp SMM4  & B & B & B       & B & B \nl
Serp SMM3  & R & N & N       & N & N \nl
B335       & B & B & B       & B & R \nl
L1157      & B & B & R       & B & B \nl

\enddata
\tablenotetext{}{R and B indicate the red and blue asymmetries, respectively, 
while N means near symmetry.  Asymmetries of CS 2--1 and H$_2$CO 
2$_{12}$--1$_{11}$ are from Mardones et al. (1997), and those of HCO$^+$ 4--3 
and 3--2 from Gregersen et al. (1997).} 
\end{deluxetable}

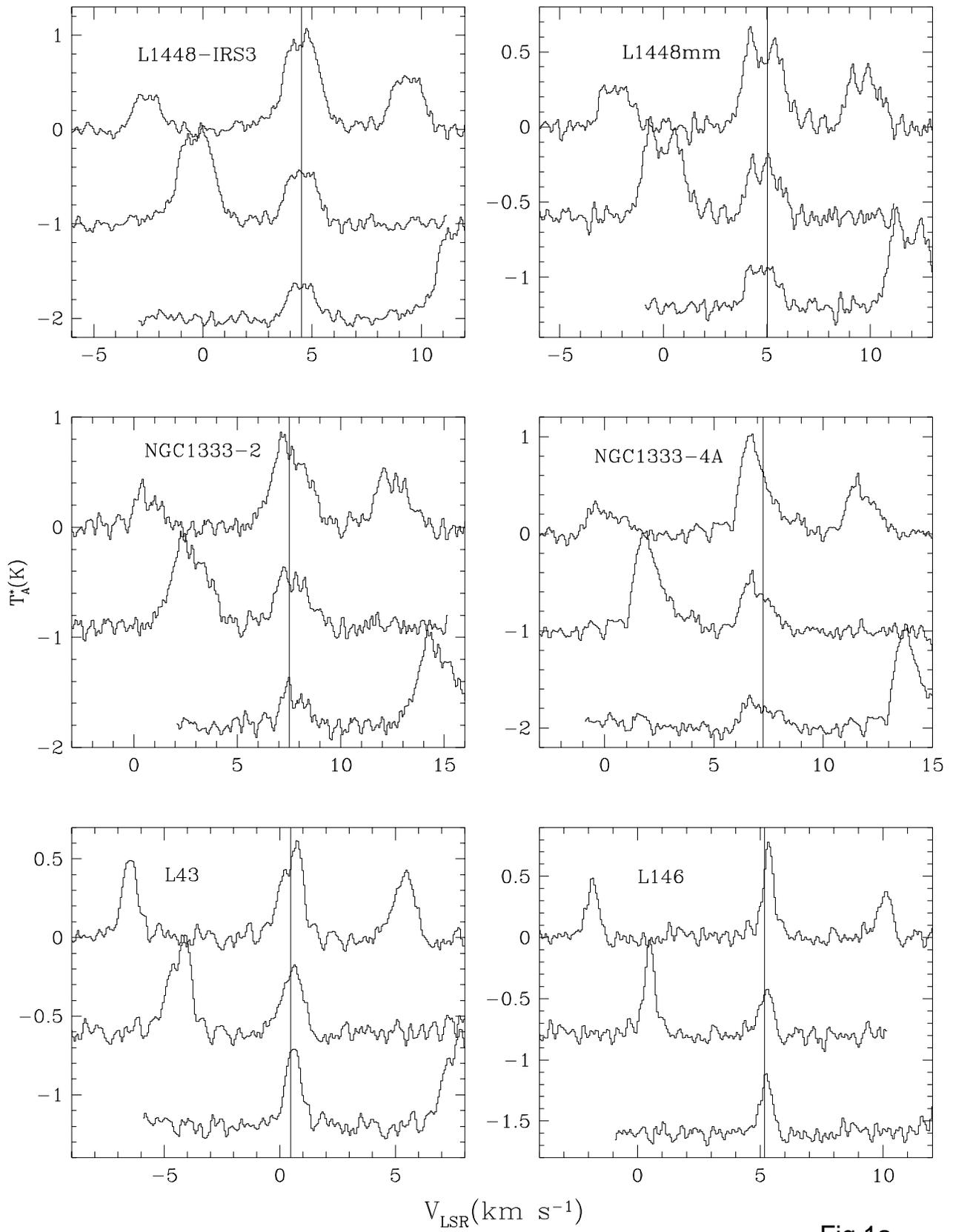
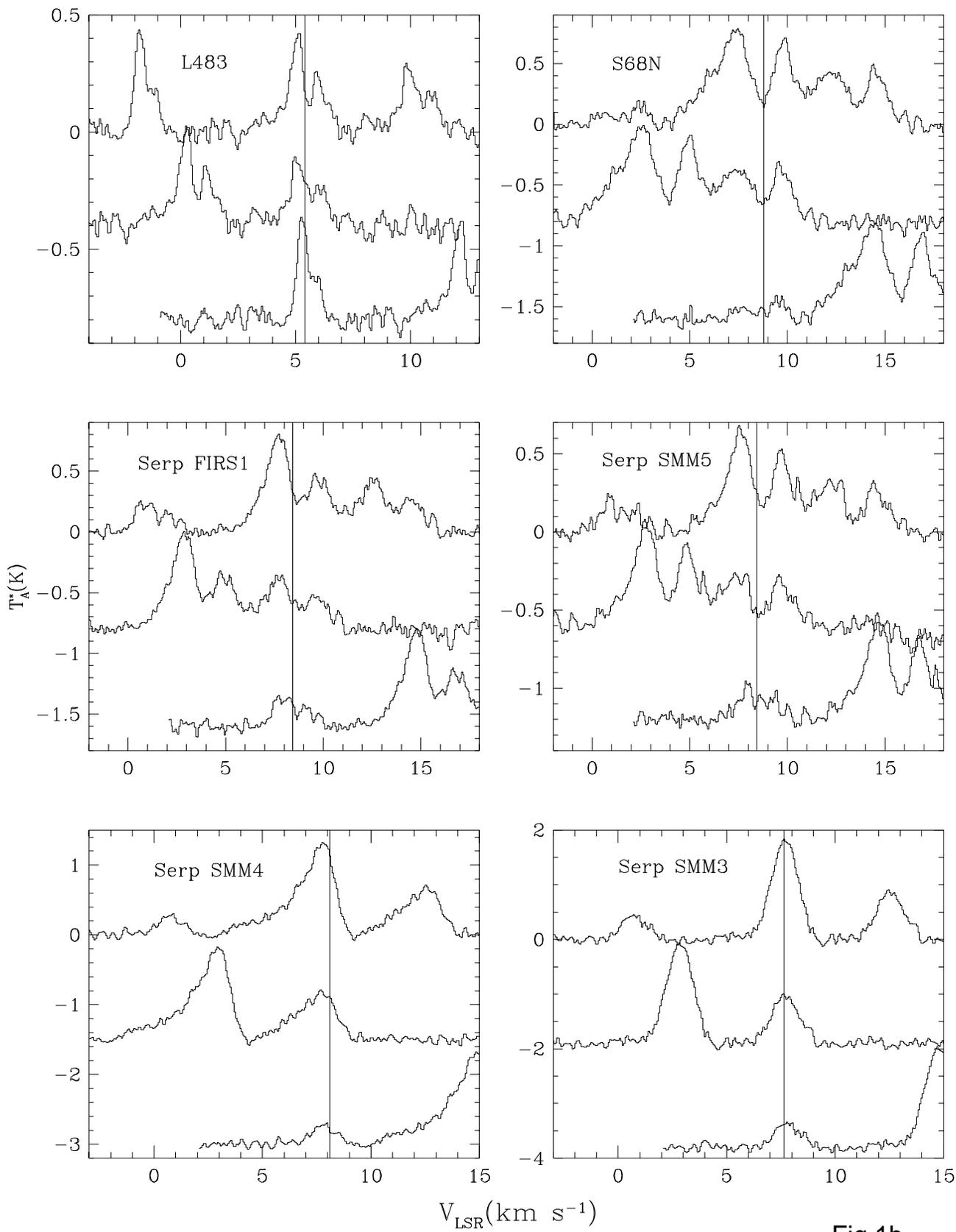
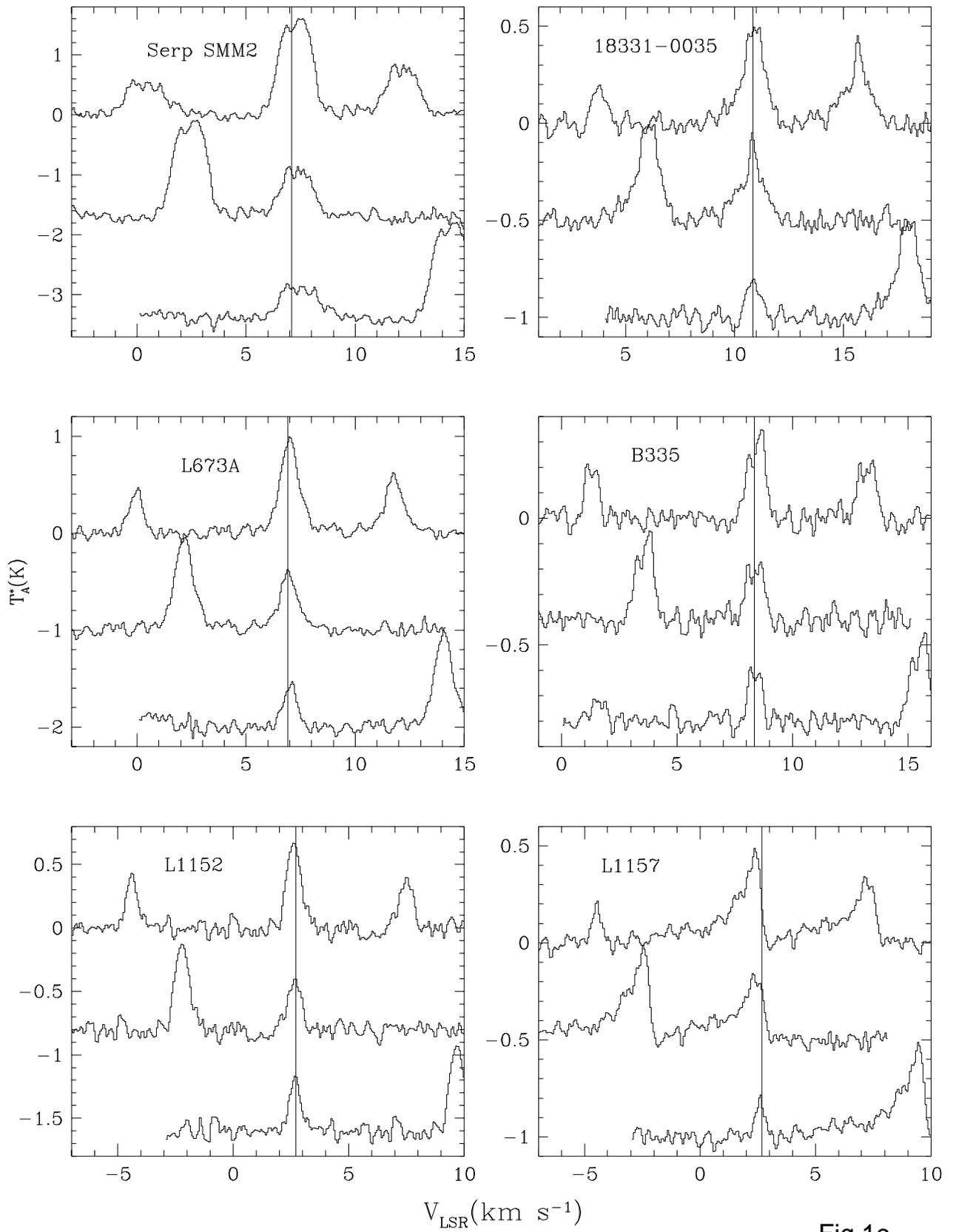
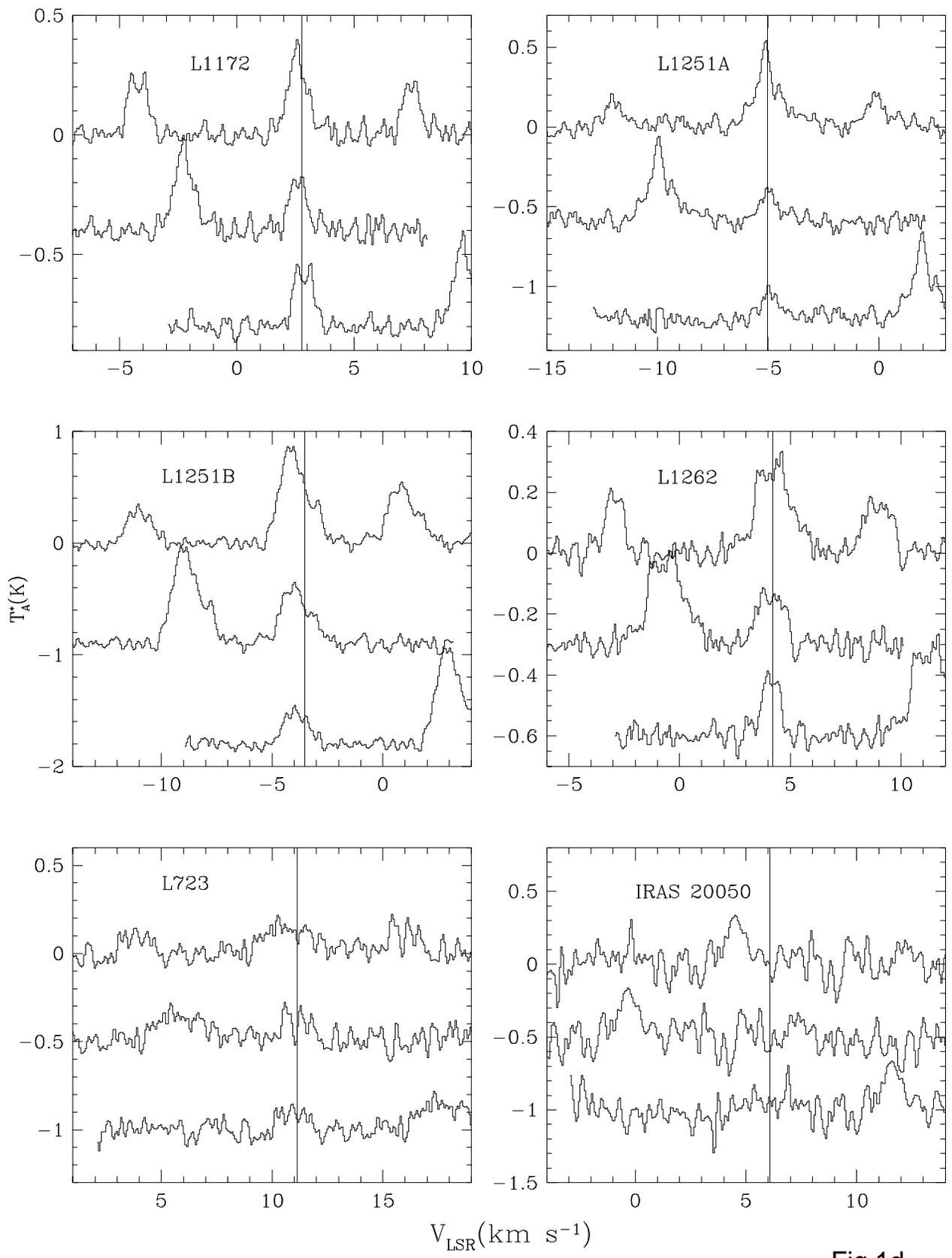
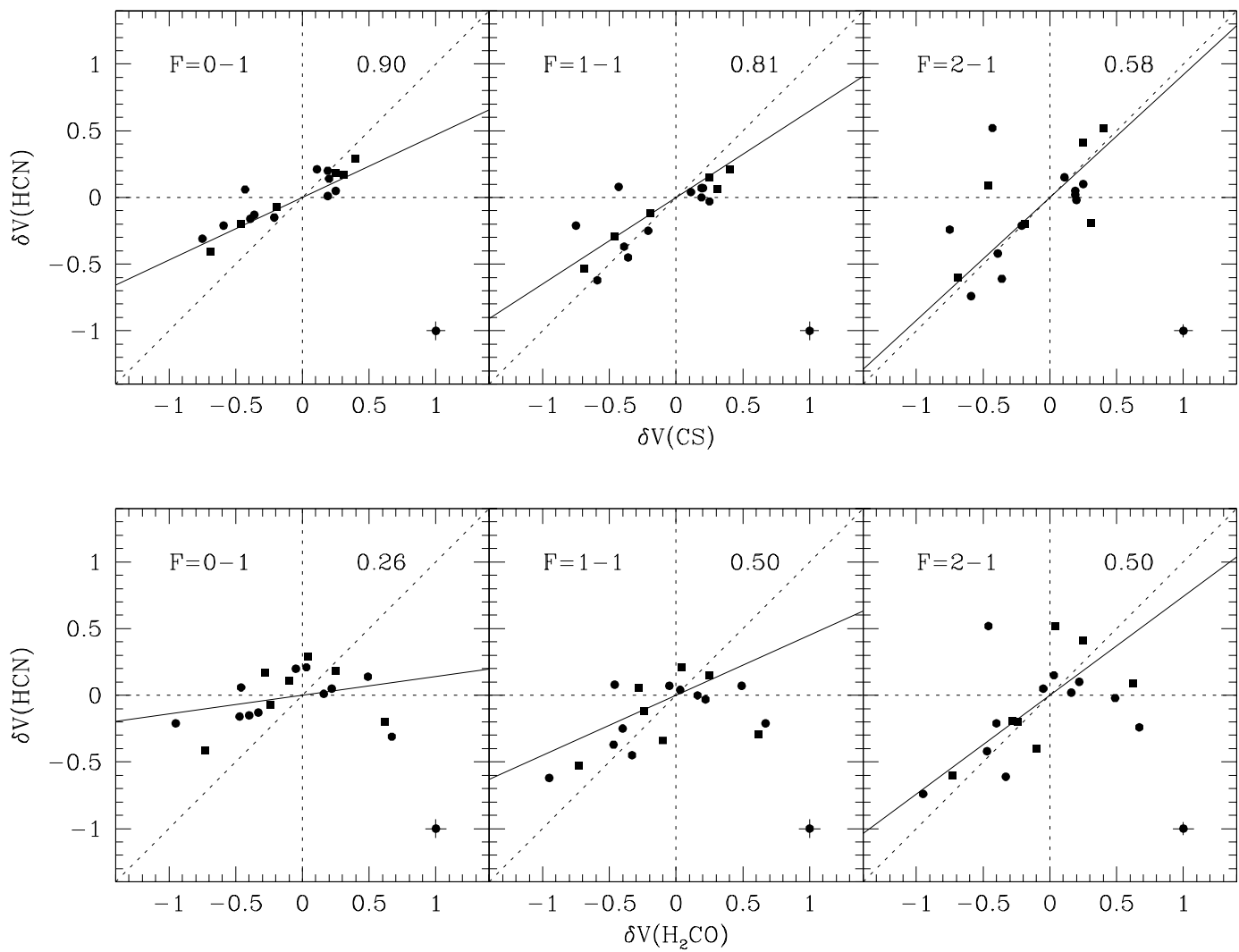
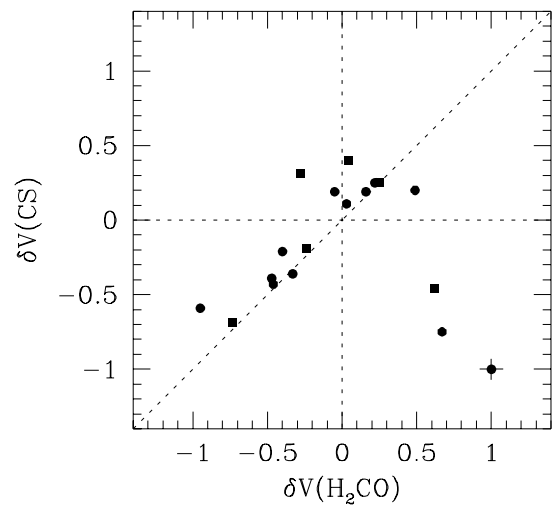
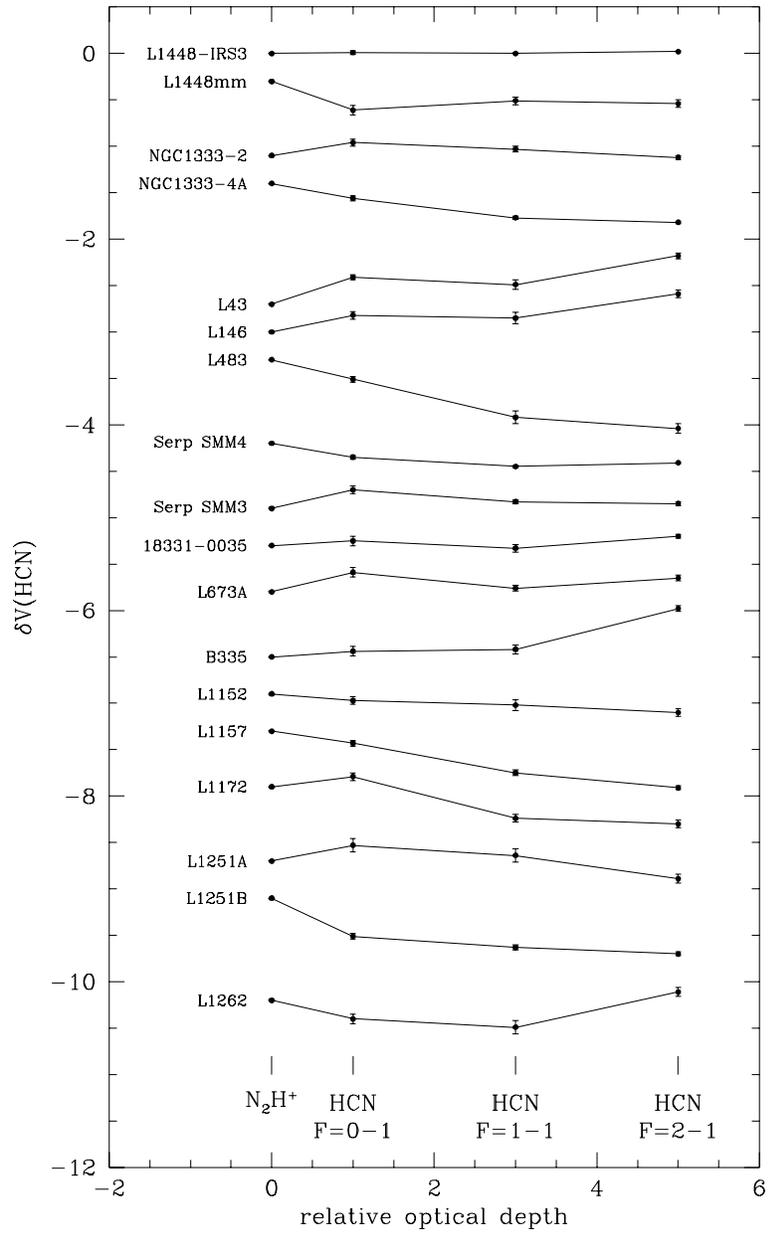
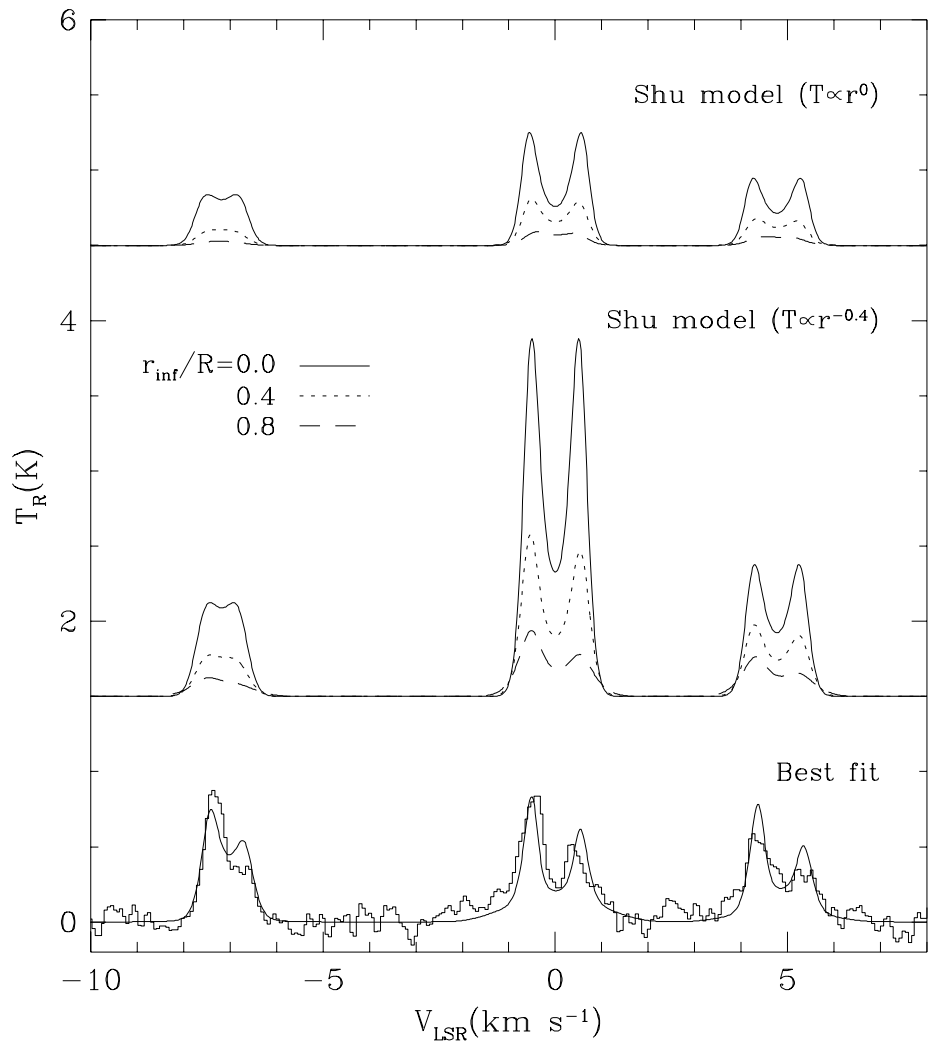
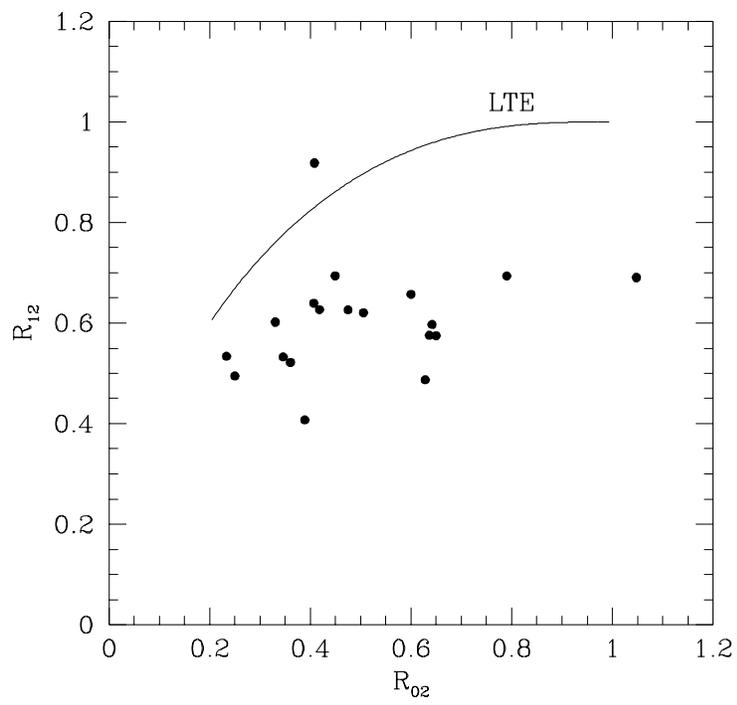
\end{document}